\documentclass[a4paper]{article}

\usepackage{amsmath, amssymb}
\usepackage[thmmarks]{ntheorem}
\usepackage[utf8]{inputenc}
\usepackage[T1]{fontenc}
\usepackage[usenames]{xcolor}
\usepackage[hyphens]{url}
\usepackage{bm}
\usepackage{todonotes}
\usepackage{booktabs}

\usepackage{microtype} %
\usepackage{fullpage}
\usepackage{algorithm}
\usepackage[noend]{algorithmic}
\usepackage{authblk}
\newcommand{\ttlpar}[1]{\smallskip\noindent{\bfseries\boldmath #1}\enspace}

\theorembodyfont{\upshape}
\theoremheaderfont{\sc}
\newtheorem{theorem}{Theorem}[section]

\theoremheaderfont{\it}
\theoremstyle{nonumberplain}
\theoremseparator{}
\theoremsymbol{\rule{1ex}{1ex}}

\newcommand{\sort}{\operatorname{sort}}
\newcommand{\scan}{\operatorname{scan}}

\newcommand{\Degree}{\operatorname{Degree}}
\newcommand{\AddEdge}{\operatorname{AddEdge}}
\newcommand{\DeleteEdge}{\operatorname{DeleteEdge}}
\newcommand{\RetrieveEdge}{\operatorname{RetrieveEdge}}

\title{Cache-Oblivious Peeling of Random Hypergraphs\thanks{Paolo
    Boldi and Sebastiano Vigna were supported by the EU-FET grant
    NADINE (GA 288956). Giuseppe Ottaviano was supported by Midas EU
    Project (318786), MaRea project (POR-FSE-2012), and
    Tiscali. Rossano Venturini was supported by the MIUR of Italy
    project PRIN ARS Technomedia 2012 and the eCloud EU Project
    (325091).}}

\author[1]{Djamal Belazzougui}
\author[2]{Paolo Boldi}
\author[3]{Giuseppe Ottaviano}
\author[4]{\\Rossano Venturini}
\author[2]{Sebastiano Vigna}
\affil[1]{Department of Computer Science\\University of Helsinki\\\texttt{djamal.belazzougui@cs.helsinki.fi}}
\affil[2]{Dipartimento di Informatica\\Università degli Studi di Milano\\\texttt{\{boldi,vigna\}@di.unimi.it}}
\affil[3]{ISTI-CNR\\Pisa\\\texttt{giuseppe.ottaviano@isti.cnr.it}}
\affil[4]{Dipartimento di Informatica\\Università di Pisa\\\texttt{rossano@di.unipi.it}}

\date{}
\begin{document}
\maketitle
\begin{abstract}
  The computation of a peeling order in a randomly generated
  hypergraph is the most time-consuming step in a number of
  constructions, such as perfect hashing schemes, random $r$-SAT
  solvers, error-correcting codes, and approximate set
  encodings. While there exists a straightforward linear time
  algorithm, its poor I/O performance makes it impractical for
  hypergraphs whose size exceeds the available internal memory.

  We show how to reduce the computation of a peeling order to a small
  number of sequential scans and sorts, and analyze its I/O complexity
  in the cache-oblivious model. The resulting algorithm requires
  $O(\sort(n))$ I/Os and $O(n \log n)$ time to peel a random
  hypergraph with $n$ edges.

  We experimentally evaluate the performance of our implementation
  of this algorithm in a real-world scenario by using the
  construction of minimal perfect hash functions (MPHF) as our test
  case: our algorithm builds a MPHF of $7.6$ billion keys in less than
  $21$ hours on a single machine.
  The resulting data structure is both more space-efficient and faster
  than that obtained with the current state-of-the-art MPHF
  construction for large-scale key sets.
\end{abstract}

\newpage

\section{Introduction}\label{sec:intro}
Hypergraphs can be used to model sets of dependencies among variables
of a system: vertices correspond to variables and edges to relations
of dependency among variables, such as equations binding variables
together. This correspondence can be used to transfer
graph-theoretical properties to solvability conditions in the original
system of dependencies.

Among these, one of the most useful is the concept of peeling
order. Given an $r$-hypergraph, a \emph{peeling order} is an order of
its edges such that each edge has a vertex of degree $1$ in the
subgraph obtained by removing the previous edges in the order. Such an
order exists if the hypergraph does not have a non-empty $2$-core,
i.e. a set of vertices that induces a subgraph whose vertices have all
degree at least $2$.

In the above interpretation, if the equations of a system are arranged
in peeling order, then each equation has at least one variable that
does not appear in any equation that comes later in the ordering,
i.e., the system becomes \emph{triangular}, so it can be easily solved
by backward substitution.
For this reason, peeling orders found application in a number of
fundamental problems, such as hash
constructions~\cite{BBPTPMMPH,BotelhoPZ13,ChCBFSL,CKRBF,CHM97,DietzfelbingerGMMPR10,MWHFPHM},
solving random instances of
$r$-SAT~\cite{DietzfelbingerGMMPR10,Molloy04,MolCRHBF}, and
the construction of error-correcting codes~\cite{Invbloom,LubyMSS01,biff}.  These
applications exploit the guarantee that if the edge sparsity $\gamma$
of a random $r$-hypergraph is larger than a certain sparsity threshold
$c_r$ (e.g., $c_3 \approx 1.221$), then with high probability the
hypergraph has an empty $2$-core \cite{MolCRHBF}.

The construction of \emph{perfect hash functions} (PHF) is probably
the most important of the aforementioned applications.  Given a set
$S$ of $n$ keys, a PHF for $S$ maps the $n$ keys onto the set of the
first $m$ natural numbers bijectively. A perfect hash function is
\emph{minimal} (MPHF) if $m=n=|S|$.  A lower bound by
Mehlhorn~\cite{Meh82} states that $n\log e \approx 1.44 n$ bits are
necessary to represent a MPHF; a matching (up to lower order terms)
upper bound is provided in~\cite{HT}, but the construction is
impractical.  Most practical approaches, instead, are based on random
$3$-hypergraphs, resulting in MPHFs that use about $2c_3n \approx
2.5n$ bits~\cite{BotelhoPZ13,CKRBF,MWHFPHM}. These solutions, which we
review in Section~\ref{sec:mwhc}, build on the MWHC technique
\cite{MWHFPHM}, whose most demanding task is in fact the computation
of a peeling order.

There is a surprisingly simple greedy algorithm to find a peeling
order when it exists, or a $2$-core when it does not: find a vertex of
degree $1$, remove (\emph{peel}) its only edge from the hypergraph,
and iterate this process until either no edges are left (in which case
the removal order is a peeling order), or all the non-isolated
vertices left have degree at least $2$ (thus forming a $2$-core). This
algorithm can be easily implemented to run in linear time and space.

MPHFs are the main ingredient in many space-efficient data structures,
such as (compressed) full-text indexes~\cite{DjamalEsa11}, monotone
MPHFs~\cite{BBPMMPH}, Bloom filter-like data structures~\cite{SODA13},
and prefix-search data structures~\cite{Weak}.

It should be clear that the applications that benefit the most from
such data structures are those involving large-scale key sets, often
orders of magnitude larger than the main memory. Unfortunately,
the standard linear-time peeling algorithm requires several tens of
bytes per key of working memory, even if the final data structure can
be stored in just a handful of bits per key. It is hence common that,
while the data structure fits in memory, such memory is not enough to
actually \emph{build} it. It is then necessary to resort to external
memory, but the poor I/O performance of the algorithm makes such an
approach impossible.

Application-specific workarounds have been devised; for example,
Botelho et al. \cite{BotelhoPZ13} proposed an algorithm (called HEM)
to build MPHFs in external memory by splitting the key set into small
buckets and computing independent MPHFs for each bucket. A first-level
index is used to find the bucket of a given key. The main drawback of
this solution is that the first-level index introduces a
non-negligible overhead in both space and lookup time; moreover, this
construction cannot be extended to applications other than hashing.

In this paper we provide the first efficient algorithm in the
\emph{cache-oblivious} model that, given a random $r$-hypergraph with
$n$ edges and $\gamma n$ vertices (with $r=O(1)$ and $\gamma>c_r$),
computes a peeling order in time $O(n\log n)$ and with $O(\sort(n))$
I/Os w.h.p., where $\sort(n)$ is the I/O complexity of sorting $n$
keys.  By applying this result we can construct (monotone) MPHFs,
static functions, and Bloom filter-like data structures in
$O(\sort(n))$ I/Os.  In our experimental evaluation, we show that the
algorithm makes it indeed possible to peel very large hypergraphs: an
MPHF for a set of $7.6$ billion keys is computed in less than $21$
hours; on the same hardware, the standard algorithm would not be able
to manage more than $2.1$ billion keys. Although we use minimal
perfect hash functions construction as our test case, results of these
experiments remain valid for all the other applications due to the
random nature of the underlying hypergraphs.

\section{Notation and tools}
\label{sec:notation}

\ttlpar{Model and assumptions}
We analyze our algorithms in the cache-oblivious
model~\cite{FrigoLPR99}. In this model, the machine has a two-level
memory hierarchy, where the fast level has an unknown size of $M$
words and a slow level of unbounded size where our data reside. We
assume that the fast level plays the role of a cache for the slow
level with an optimal replacement strategy where the transfers
(a.k.a.~I/Os) between the two levels are done in blocks of an unknown
size of $B\leq M$ words; the I/O cost of an algorithm is the total number
of such block transfers.
\emph{Scanning} and \emph{sorting} are two fundamental building blocks
in the design of cache-oblivious algorithms~\cite{FrigoLPR99}: under
the tall-cache assumption \cite{BrodalF03}, given an array of $N$
contiguous items the I/Os required for scanning and sorting are
\begin{equation*}
  \scan(N)=O\left(1+\frac NB\right) \text{ I/Os} \quad\text{and}\quad
  \sort(N)=O\left(\frac NB\log_{M/B}\frac NB\right) \text{.}
\end{equation*}

\ttlpar{Hypergraphs}%
An $r$-hypergraph on a vertex set $V$ is a subset $E$ of ${V \choose r}$, the
set of subsets of $V$ of cardinality $r$. An element of $E$ is called
an \emph{edge}. We call an ordered $r$-tuple from $V$ an
\emph{oriented edge}; if $e$ is an edge, an oriented edge whose
vertices are those in $e$ is called an \emph{orientation} of $e$.
From now on we will focus on $3$-hypergraphs; generalization to
arbitrary $r$ is straightforward. We define \emph{valid} orientations
those oriented edges $(v_0, v_1, v_2)$ where $v_1 < v_2$ (for
arbitrary $r$, $v_1 < \dots < v_{r-1}$). Then for each edge there are
$6$ orientations, but only $3$ valid orientations ($r!$ orientations
of which $r$ are valid).

We say that a valid oriented edge $(v_0, v_1, v_2)$ is the $i$-th
orientation if $v_0$ is the $i$-th smallest among the three; in
particular, the $0$-th orientation is the \emph{canonical}
orientation. Edges correspond bijectively with their canonical
orientations. Furthermore, valid orientations can be mapped
bijectively to pairs $(e, v)$ where $e$ is an edge and $v$ a vertex
contained in $e$, simply by the correspondence $(v_0, v_1, v_2)
\mapsto (\{v_0, v_1, v_2\}, v_0)$. In the following all the
orientations are assumed to be valid, so we will use the term
\emph{orientation} to mean \emph{valid orientation}.

\section{The Majewski--Wormald--Havas--Czech technique}
\label{sec:mwhc}

Majewski et al.~\cite{MWHFPHM} proposed a technique (MWHC) to compute
an \emph{order-preserving minimal perfect hash function}, that is, a
function mapping a set of keys $S$ in some specified way into
$[|S|]$. The technique actually makes it possible to store succinctly
any function $f:S\to [\sigma]$, for arbitrary $\sigma$.  In this
section we briefly describe their construction.

First, we choose three random\footnote{Like most MWHC implementations,
  in our experiments we use a Jenkins hash function with a $64$-bit
  seed in place of a fully random hash function.} hash functions $h_0,
h_1, h_2: S\to [\gamma n]$ and generate a
3-hypergraph\footnote{Although the technique works for
  $r$-hypergraphs, $r=3$ provides the lowest space
  usage~\cite{MolCRHBF}.}  with $\gamma n$ vertices, where $\gamma$ is
a constant above the \emph{critical threshold} $c_3$~\cite{MolCRHBF},
by mapping each key $x$ to the edge $\{h_0(x), h_1(x),h_2(x)\}$.  The
goal is to find an array $u$ of $\gamma n$ integers in $[\sigma]$ such
that for each key $x$ one has $f(x) = u_{h_0(x)} + u_{h_1(x)} +
u_{h_2(x)} \mod \sigma$. This yields a linear system with $n$
equations and $\gamma n$ variables $u_i$; if the associated hypergraph
is peelable, it is easy to solve the system. Since $\gamma$ is larger
than the critical threshold, the algorithm succeeds with probability
$1 - o(1)$ as $n \to \infty$ \cite{MolCRHBF}.

By storing such values $u_i$, each requiring $\lceil \log \sigma
\rceil$ bits, plus the three hash functions, we will be able to
recover $f(x)$. Overall, the space required will be $\lceil \log
\sigma \rceil \gamma n$ bits, which can be reduced to $\lceil \log
\sigma \rceil n+\gamma n+o(n)$ using a ranking structure~\cite{Jac89}.
This technique can be easily extended to construct MPHFs: we define
the function $f\colon S \to [3]$ as $x \mapsto i$ where $h_i(x)$ is
a degree-$1$ vertex when the edge corresponding to $x$ is peeled; it
is then easy to see that $h_{f(x)}(x): S \to [\gamma n]$ is a PHF. The
function can be again made minimal by adding a ranking structure on
the vector $u$~\cite{BotelhoPZ13}.

As noted in the introduction, the peeling procedure needed to solve
the linear system can be performed in linear time using a greedy
algorithm (referred to as \emph{standard linear-time
  peeling}). However, this procedure requires random access to several
integers per key, needed for bookkeeping; moreover, since the graph is
random, the visit order is close to random.
As a consequence, if the key set is so large that it is necessary to
spill to the disk part of the working data structures, the I/O volume
slows down the algorithm to unacceptable rates.

\ttlpar{Practical workarounds (HEM)}%
Botelho et al. \cite{BotelhoPZ13} proposed a practical external-memory solution:
they replace each key with a \emph{signature}
of $\Theta(\log n)$ bits computed with a random hash function, so that
no collision occurs. The signatures are then sorted and divided into
small buckets based on their most significant bits, and a separate
MPHF is computed for each bucket with the approach described
above. The representations of the bucket functions are then
concatenated into a single array and their offsets stored in a
separate vector.

The construction algorithm only requires to sort the signatures (which
can be done efficiently in external memory) and to scan the resulting
array to compute the bucket functions; hence, it is extremely
scalable.  The extra indirection needed to address the blocks causes
however the resulting data structure to be both \emph{slower} and
\emph{larger} than one obtained by computing a single function on the
whole key set. In their experiments with a practical version of the
construction, named HEM, the authors report that the resulting data
structure is $21\%$ larger than the one built with plain MWHC, and
lookups are $30\text{--}50\%$ slower. A similar overhead was confirmed
in our experiments, which are discussed in
Section~\ref{sec:experiments}.

\section{Cache-oblivious peeling}
\label{sec:copeel}

In this section we describe a cache-oblivious algorithm to peel an
$r$-hypergraph. We describe the algorithm for $3$-hypergraphs, but it
is easy to generalize it to arbitrary $r$.

\subsection{Maintaining incidence lists}
\label{sec:xortrick}

In order to represent the hypergraph throughout the execution of the
algorithm, we need a data structure to store the \emph{incidence list}
of every vertex $v_0$, i.e., the list $L_{v_0} = \{(v_0, v_1^{0},
v_2^{0}),\allowbreak \dots,\allowbreak (v_0, v_1^{d-1}, v_2^{d-1})\}$ of \emph{valid}
oriented edges whose first vertex is $v_0$.
To realize the peeling algorithm, it is sufficient to implement the
following operations on the lists.
\begin{itemize}
\item $\Degree(L_{v_0})$ returns the number of edges $d$ in the
  incidence list of $v_0$;
\item $\AddEdge(L_{v_0}, e)$ adds the edge $e$ to the incidence list
  of $v_0$;
\item $\DeleteEdge(L_{v_0}, e)$ deletes the edge $e$ from the
  incidence list of $v_0$;
\item $\RetrieveEdge(L_{v_0})$ returns the only edge in the list if
  $\Degree(L_{v_0}) = 1$. %
\end{itemize}

For all the operations above, it is assumed that the edge is
given through a \emph{valid} orientation. Under this set of operations, the data structure does
not need to \emph{store} the actual list of edges: it is sufficient to
store a tuple $(v_0, d, \tilde v_1, \tilde v_2)$, where $d$ is the
number of edges, $\tilde v_1 = \bigoplus_{j < d} v_1^j$, and $\tilde
v_2 = \bigoplus_{j < d} v_2^j$, that is, all the vertices of the list
in the same position are XORed together.

The operations $\AddEdge$ and $\DeleteEdge$ on an edge $(v_0, v_1',
v_2')$ simply XOR $v_1'$ into $\tilde v_1$ and $v_2'$ into $\tilde
v_2$, and respectively increment or decrement $d$. Since all the
edges are assumed valid (i.e., it holds that $v_1'<v_2'$) these
operations maintain the invariant. When $d = 1$, clearly $\tilde v_1 =
v_1$ and $\tilde v_2 = v_2$ where $(v_0, v_1, v_2)$ is the only edge
in $L_{v_0}$, so it can be returned by $\RetrieveEdge$. If necessary, the data
structure can be trivially extended to \emph{labeled} edges
$(v_0, v_1, v_2, \ell)$ by XORing together the labels $\ell$ into a
new field $\tilde \ell$.

We call this data structure \emph{packed incidence list}, and we refer
to this technique as the \emph{XOR trick}. The advantage with respect
to maintaining an explicit list, besides the obvious space savings, is
that it is sufficient to maintain a single fixed-size record per
vertex, regardless of the number of incident edges. This will make the
peeling algorithm in the next section substantially simpler and
faster.  The same trick can be applied to the standard linear-time
algorithm, replacing the linked lists traditionally used. As we will
see in Section~\ref{sec:experiments}, the improvements are significant
in both working space and running time.

\subsection{Layered peeling}
\label{sec:copeeling}

The peeling procedure we present
is an adaptation of the CORE procedure presented by
Molloy~\cite{MolCRHBF}. The basic idea is to proceed in rounds: at
each round, all the vertices of degree $1$ are removed, and then the
next round is performed on the induced subgraph, until either a
$2$-core is left, or the graph is empty. In the latter case, the
algorithm partitions the edges into a sequence of \emph{layers}, one per round, by
defining each layer as the set of edges removed in its round. It is
easy to see that by concatenating the layers the resulting edge order
is a peeling order, regardless of the order within each layer.

The layered peeling process terminates in a small number of rounds:
Jiang et al. \cite{JMTPPA} proved that if the hypergraph is generated
randomly with a sparsity above the peeling threshold, then with high
probability the number of rounds is bounded by $O(\log \log
n)$. Moreover, the fraction of vertices remaining in each round
decreases double-exponentially.  In the following we show how to
implement the algorithm in an I/O-efficient way by putting special
care in the hypergraph representation and the update step.

\ttlpar{Hypergraph representation}%
At each round $i$, the hypergraph is represented by a list $E_i$ of
tuples $(v_0, d, \tilde v_1, \tilde v_2)$ as described in
Section~\ref{sec:xortrick}; each tuple represents the incidence list
of $v_0$. Each list $E_i$ is sorted by $v_0$. Note that each edge $e =
\{v_0, v_1, v_2\}$ needs to be in the incidence list of all its
vertices; hence, all the three orientations of $e$ are present in the
list $E_i$.

\ttlpar{Construction of $E_0$}%
To construct $E_0$, the edge list for the first round, we put together
in a list all the valid orientations of all the edges in the
hypergraph. The list is then sorted by $v_0$, and from the sorted list
we can construct the sorted list of incidence lists $E_0$: after
grouping the oriented edges by $v_0$, we start with the empty
packed incidence list $(v_0, 0, 0, 0)$ and, after performing $\AddEdge$ with all the
edges in the group, we append it to $E_0$. The I/O complexity is $O(\sort(n) + \scan(n)) =
O(\sort(n))$.

\ttlpar{Round update}%
At the beginning of each round we are given the list $E_i$ of edges
that are alive at round $i$, and we produce $E_{i + 1}$. We first scan
$E_i$ to find all the tuples $L$ such that $\Degree(L) = 1$; for each tuple,
we perform $\RetrieveEdge$ and put the edge in a list $D_i$, which
represents all the edges to be removed in the current round $i$. The
same edge may occur multiple times in $D_i$ under different
orientations (if more than one of its vertices have degree $1$ in the
current round); to remove the duplicates, we sort the oriented edges
by their canonical orientation, keep one orientation for each edge,
and store them in a list $P_i$.

Now we need to remove the edges from the hypergraph. To do so, we
generate a \emph{degree update list} $U_i$ that contains all the three
orientations for each edge in $P_i$, and sort $U_i$ by $v_0$. Since
both $E_i$ and $U_i$ are sorted by $v_0$, we can scan them both
simultaneously joining them by $v_0$; for each tuple $L_{v_0}$ in
$E_i$, if no oriented edge starting with $v_0$ is in $U_i$ the tuple
is copied to $E_{i + 1}$, otherwise for each such oriented edge $e$,
$\DeleteEdge(L_{v_0}, e)$ is called to obtain a new $L_{v_0}'$ which
is written to $E_{i + 1}$ if non-empty. Note that $E_{i + 1}$ remains
sorted by $v_0$.

For each round, we scan $E_i$ twice and $U_i$ once, and sort $D_i$ and
$U_i$. The number of I/Os is then $2\cdot \scan(|E_i|) + \scan(|U_i|)
+ \sort(|D_i|) + \sort(|U_i|)$. Summing over all rounds, we have
$\sum_i (2\cdot \scan(|E_i|) + \sort(|D_i|) + \sort(|U_i|) +
\scan(|U_i|)) = O(\sort(n))$ because each edge belongs to at most
three lists $D_i$ and three lists $U_i$. Since the fraction of
vertices remaining at each round decreases doubly exponentially and,
thanks to the XOR trick, $E_i$ has exactly a tuple for each vertex
alive in the $i$-th round, the cost of scanning the lists $E_i$ sums
up to $O(\scan(n))$ I/Os. Hence, overall the algorithm takes $O(n \log
n)$ time and $O(\sort(n))$ I/Os.

We summarize the result in the following theorem.
\begin{theorem}
  A peeling order of a random $r$-hypergraph with $n$ edges and
  $\gamma n$ vertices with constant $r$ and $\gamma>c_r$, can be
  computed in the cache-oblivious model in time $O(n\log n)$ and with
  $O(\sort(n))$ I/Os with high probability.
\end{theorem}

\subsection{Implementation details}

We report here the most important optimizations we used in our
implementation. The source code used in the experiments is available
at \url{https://github.com/ot/emphf} for the reader interested in
further implementation details and in replicating the measurements.

\ttlpar{File I/O}%
Instead of managing file I/Os directly, we use a memory-mapped file by
employing a C++ allocator that creates a file-backed area of
memory. This way we can use the standard STL containers such as
\verb|std::vector| as if they resided in internal memory. We use
\verb|madvise| to instruct the kernel to optimize the mapped region
for sequential access.  We use the \verb|madvise| system call with the
parameter \verb|MADV_SEQUENTIAL| on the memory-mapped region to
instruct the kernel to optimize for sequential access.

\ttlpar{Sorting}%
Our sorting implementation performs two steps: in the first step we
divide the domain of the values into $k$ evenly spaced buckets,
scanning the array to find the number of values that belong in each
bucket, and then moving each value to its own bucket. In the second
step, each bucket is sorted using \verb|sort| of the C++ standard
library. The number of buckets is chosen so that with very high
probability each bucket fits in internal memory; since the graph is
random, its edges are uniformly distributed, which makes uniform
bucketing balanced with high probability. To distribute the values into the $k$ buckets, we
use a buffer of size $T$ for each bucket; when the buffer is full, it
is flushed to disk.
Note that this algorithm is technically not cache-oblivious, since it
works as long as the available memory $M$ is at least $kT$; choosing
$k$ to be $\Theta(S/M)$, where $S$ is the size of the data to be
sorted, requires that $M$ be $\Omega(\sqrt{TS})$.  In our
implementation we use $T \approx 1$MiB, thus for example $M = 1$GiB is
sufficient to sort ${\approx}1$TiB of data. When this condition holds,
the algorithm performs just three scans of the array and it is
extremely efficient in practice. Furthermore, contrary to existing
cache-oblivious sorting implementations, it is \emph{in-place}, using
no extra disk space.

\ttlpar{Reusing memory}%
The algorithm as described in Section~\ref{sec:copeeling} uses a
different list $E_i$ for each round. Since tuples are appended to
$E_{i+1}$ at a slower pace than they are read from $E_i$, we can reuse
the same array. A similar trick can be applied to $D_i$ and $U_i$.
Overall, we need to allocate just one array of $\gamma n$ packed
incidence lists, and one for the $3n$ oriented edges.

\ttlpar{Lists compression}%
Reducing the size of the on-disk data structures can significantly
improve I/O efficiency, and hence the running time of the
algorithm. The two data structures that take nearly all the space are
the lists of packed incidence lists $E_i$ and the lists of edges
$P_i$. Since the lists are read and written sequentially, we can
\emph{(de)compress} them on the fly.

Recall that the elements of $E_i$ are tuples of the form $(v_0, d,
\tilde v_1, \tilde v_2)$ sorted by $v_0$. The first components $v_0$
of these tuples are gap-encoded with Elias $\gamma$ codes.  The
overall size of the encoding is $\sum_{k=1} ^{|E_i|} (2\lfloor \log
g_k \rfloor + 1)$ bits, where $g_k$ is the $k$-th gap.
\footnote{Elias Gamma code \cite{Elias75} uses $2\lfloor \log j
  \rfloor + 1$ bits to encode any integer $j \geq 1$.}.  Since the
gaps sum up to $\gamma n$, by Jensen's inequality the sum is maximized
when the gaps $g_k$ are all equal to $\frac{\gamma n}{|E_i|}$ giving a
space bound of $2 |E_i| (\log \frac{\gamma n}{|E_i|} + 1)$ bits.
Furthermore, this space bound is always at most $2\gamma n$ bits
because it is maximized when $E_i$ has size $\gamma n/2$.  The degrees
$d$ are encoded instead with unary codes; since the sum of the degrees
is $3n_i$, where $n_i$ is the number of edges alive in round $i$, the
overall size of their encoding is always upper bounded by $3n$
bits. The other two components, as well as the nodes in $P_i$, are
represented with a fixed-length encoding using $\lceil \log \gamma n
\rceil$ bits each. With $\gamma = 1.23$, and thanks to the memory
reusing technique described above, the overall disk usage is
approximately $(5.46 + 11.46 \lceil \log \gamma n \rceil)n$ bits. On
our largest inputs, using compression instead of plain 64-bit words
makes the overall algorithm run about $2.5$ times faster.

\ttlpar{Exploiting the tripartition}%
Many MWHC-based implementations, when generating the $r$-hypergraph
edges $\{h_0(x), \dots, h_{r-1}(x)\}$, use random hash functions $h_i$
with codomain $[i |V| / r, (i+1)|V| / r)$ instead of $[0, |V|)$, thus
yielding a $r$-partite $r$-hypergraph. The main advantage is that by
construction $h_i(x) \neq h_j(x)$ for $i \neq j$, so the process
cannot generate hypergraphs with degenerate edges; this reduces
considerably the number of trials needed to find a peelable hypergraph
(in practice, just one trial is sufficient). Botelho et
al. \cite{Botelho2012314} proved that hypergraphs obtained with this
process have the same peeling threshold as uniformly random
hypergraphs. Jiang et al. \cite{JMTPPA} proved that the
bound on the number of rounds of the layered peeling process also
holds for random $r$-partite $r$-hypergraphs, so we can adopt this
approach as well.

An additional advantage of the $r$-partition is that the first vertex
of any $0$-orientation is smaller than the first vertex of any
$1$-orientation, and so on; in general, if $(u_0, \dots, u_{r-1})$
is an $i$-orientation, $(v_0, \dots, v_{r-1})$ is a
$j$-orientation, and $i < j$, then $u_0 < v_0$. We exploit this in
our algorithm in the construction of $E_0$: since our graph is
$3$-partite, instead of creating a list with every valid orientation
of each edge and then sorting it by $v_0$, we create a list with just
the $0$-orientations, sort it by $v_0$, and append the obtained packed
incidence lists to $E_0$. Then we go through the sorted list, switch
all the oriented edges to $1$-orientation, and repeat the
process. The same is done for the $2$-orientations.

Thanks to this optimization the amount of memory required in the first
step of the algorithm, which is the most I/O intensive, is reduced to
one third.

\ttlpar{Avoiding backward scans}%
For MWHC-based functions construction, the final phase that assigns
the $u_i$s needs to scan the edges in \emph{reverse} peeling
order. Unfortunately, operating systems and disks are highly optimized
for \emph{forward} reading, by performing an aggressive
lookahead. However, as we noted in Section~\ref{sec:copeeling}, the
ordering of the edges \emph{within} the layers is irrelevant; thus it
is sufficient to scan the layers in reverse order, but each layer may
be safely scanned forward. The number of forward scans is then bounded
by the number of rounds, which is negligible. The performance
improvement of the assignment phase with respect to reading the array
backwards is almost ten-fold.

\section{Experimental analysis}
\label{sec:experiments}

Although our code can be easily extended to construct \emph{any}
static function, to evaluate experimentally the performance of the
peeling algorithm we tested it on the task of constructing a minimal
perfect hash function, as discussed in Section~\ref{sec:mwhc}.  In
this task, the peeling process largely dominates the running
time.

\ttlpar{Testing details} The tests of MPHF construction were performed
on an Intel Xeon i7 E5520 (Nehalem) at 2.27GHz with 32GiB of RAM,
running Linux~3.5.0 x86-64. The storage device is a 3TB Western
Digital WD30EFRX hard drive. Before running each test, the kernel page
cache was cleared to ensure that all the data were read from disk. The
experiments were written in C++11 and compiled with g++ 4.8.1 at
\texttt{-O3}.

We tested the following algorithms.

\begin{itemize}
\item \textsf{Cache-Oblivious}: The cache-oblivious algorithm
  described in Section~\ref{sec:copeel}.
\item \textsf{Standard+XOR}: The standard linear-time peeling
  implemented using the packed incidence list, with the purpose of
  evaluating the impact of the XOR-trick by itself.
\item \textsf{cmph}: A publicly available, widely used and optimized
  library for minimal perfect hashing\footnote{We used
    \textsf{cmph~2.0}, available at
    \url{http://cmph.sourceforge.net/}.}, implementing the same
  MWHC-based MPHF construction with the standard in-memory peeling
  algorithm.

\end{itemize}

\ttlpar{Datasets}%
We tested the above algorithms on the following datasets.

\begin{itemize}
\item \textsf{URLs}: a set of ${\approx}4.8$ billion URLs from the
  ClueWeb09 dataset\footnote{Downloaded from
    \url{http://lemurproject.org/clueweb09/}.} (average string
  length ${\approx}67$ bytes, summing up to ${\approx}304$GiB);
\item \textsf{ngrams}: a set of ${\approx}7.6$ billion
  $\{1,2,3\}$-grams obtained from the Google Books Ngrams English
  dataset\footnote{Downloaded from 
    \url{http://storage.googleapis.com/books/ngrams/books/datasetsv2.html}.}
  (average string length ${\approx}23$ bytes, summing up to
  ${\approx}168$GiB).
\end{itemize}

Since the strings are hashed in the first place, the nature of the
data is fairly irrelevant: the only aspect that may be relevant is the
average string length (that affects the time to load the input from
disk). In fact tests on randomly generated data produced the same
results.

\begin{figure}[htb]
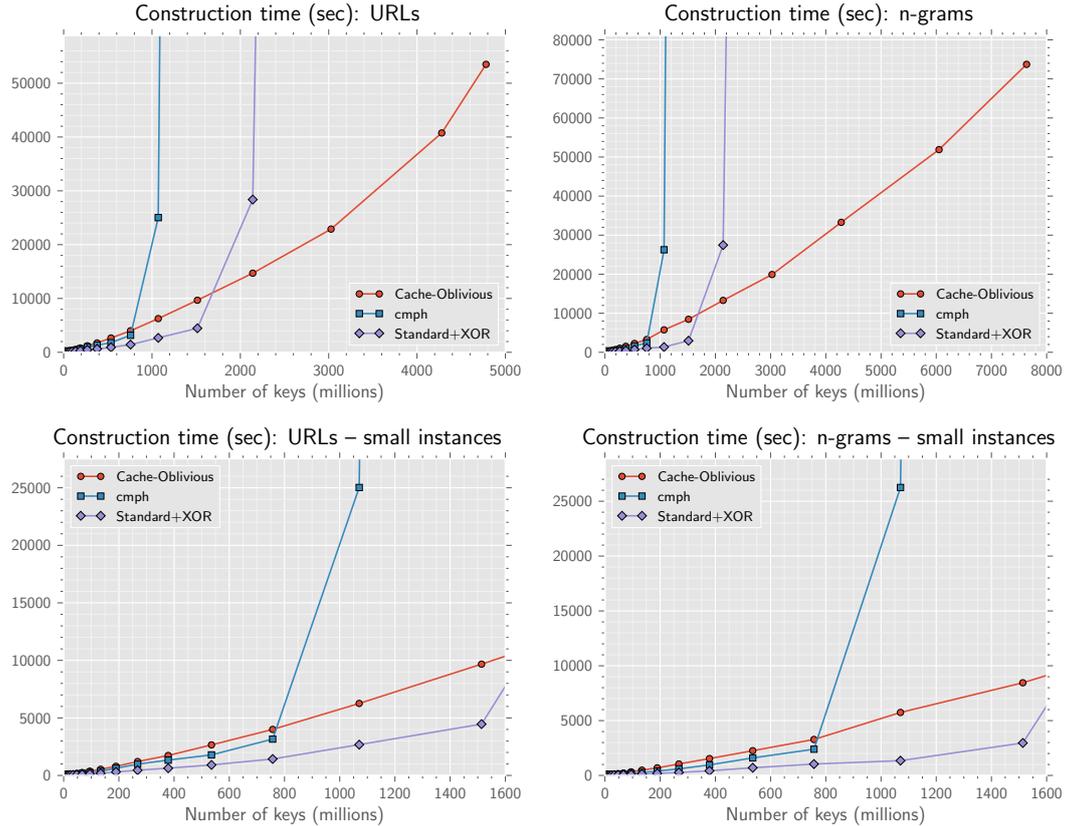

  \centering
  \includegraphics[width=0.44\linewidth]{{{figures/build_times.clueweb_big}}}
  \includegraphics[width=0.44\linewidth]{{{figures/build_times.ngrams_big}}}

  \includegraphics[width=0.44\linewidth]{{{figures/build_times.small.clueweb_big}}}
  \includegraphics[width=0.44\linewidth]{{{figures/build_times.small.ngrams_big}}}
  \caption{Above: construction times on the two datasets. Below:
    close-up for $n$ up to $1.6 \cdot 10^9$ keys.}
\label{fig:buildtimes}
\end{figure}

\ttlpar{Experimental results}%
The running time of the algorithms as the number of keys increases is
plotted in Figure~\ref{fig:buildtimes}; to evaluate the performance in
the regime where the working space fits in main memory, the figure
also shows an enlarged version of the first part of the plot.

The first interesting observation is that the cache-oblivious
algorithm performs almost as well as \textsf{cpmh}, with
\textsf{Cache-Oblivious} being slightly slower because it has to
perform file I/O even when the working space would fit in memory.

We can also see that the XOR trick pays off, as shown by the
performance of \textsf{Standard+XOR}, which is up to $3$ times faster
than \textsf{cmph}, and the smaller space usage enables to process up
to almost twice the number of keys for the given memory budget. Both
non-external algorithms, though, cease to be useful as soon as the
available memory gets exhausted: the machine, then, starts to thrash
because of the random patterns of access to the swap. In fact, we had
to kill the processes after $48$ hours. Actually, one can make a quite
precise estimate of when this is going to happen: \textsf{cpmh}
occupies $34.62$ bytes/key, as estimated by the authors, whereas
\textsf{Standard+XOR} occupies about $26.76$ bytes/key, and these
figures almost exactly justify the two points where the construction
times slow down and then explode.  On the other hand,
\textsf{Cache-Oblivious} scales well with the input size, exhibiting
eventually almost linear performance in our larger input
\textsf{ngrams}, while remaining competitive even on small key sets.

\ttlpar{Comparison with HEM}%
Finally, we compare our algorithm with HEM~\cite{BotelhoPZ13}.
Recall that their technique consists in splitting the set of keys into
several buckets and building a separate MPHF for every bucket; at
query time, a first-level index is used to drive the query to the
correct bucket. Choosing a sufficiently small size for the buckets
allows the use of a standard internal memory algorithm to construct
the bucket MPHF.  Although technically not a peeling algorithm, this
external-memory solution is simple and elegant.

To make a fair comparison, we re-implemented the HEM algorithm using
our sort implementation for the initial bucketing, and the
\textsf{Standard+XOR} algorithm to build the bucket MPHFs. The
signature function is the same 96-bit hash function used in
\cite{BotelhoPZ13} (which suffice for sets of up to $2^{48}$ keys),
but we employed $64$-bit bucket offsets in place of $32$-bit, since our
key sets are larger than $2^{32}/\gamma$.

\begin{figure}[htbp]
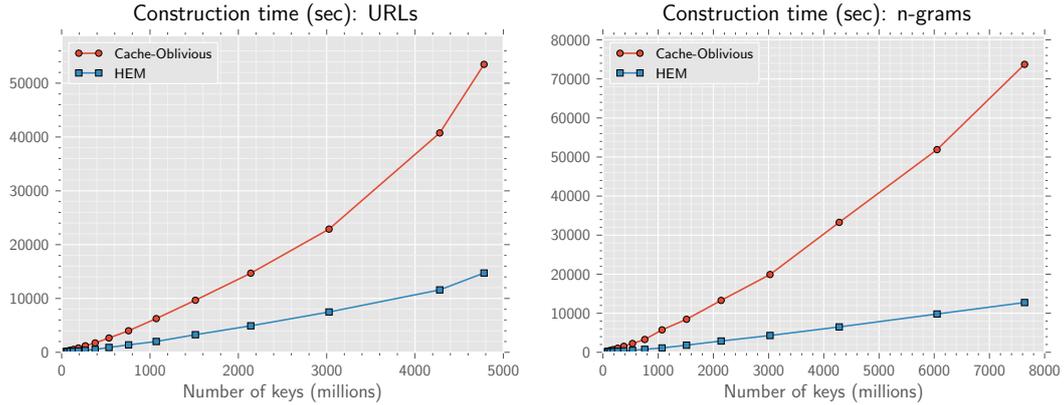

  \centering
  \includegraphics[width=0.44\linewidth]{{{figures/build_times.hem.clueweb_big}}}
  \includegraphics[width=0.44\linewidth]{{{figures/build_times.hem.ngrams_big}}}
  \caption{Construction time with the Cache-Oblivious algorithm and
    HEM~\cite{BotelhoPZ13}.}
\label{fig:buildtimeshem}
\end{figure}

The result, as shown in Figure~\ref{fig:buildtimeshem}, is a
construction time between $2$ and $6$ times smaller than
\textsf{Cache-Oblivious}.  However, this efficiency has a cost in term
of lookup time (because of the double indirection) and size (because
of the extra space needed for the first-level index). Since, in most
applications, MPHFs are built rarely and queried frequently, the
shorter construction time may not be worth the increase in space and
query time.

Indeed, as shown in Table~\ref{tab:space}, the space loss is $17\%$ to
$27\%$. The variability in space overhead is due to the fact that in
HEM the number of buckets must be a power of $2$, hence the actual
average bucket size can vary by a factor of $2$ depending on the
number of keys. We also include the space taken by \textsf{cmph} on
the largest inputs we were able to construct in-memory. Despite using
the same data structure as our implementation of \textsf{MWHC}, its
space occupancy is slightly larger because it uses denser ranking
tables.

\begin{table*}[htbp]
  { \small
    \centering
    \begin{tabular}{@{}lrrrr@{}}\toprule
      & \multicolumn{2}{c}{\textsf{URLs}} %
      & \multicolumn{2}{c}{\textsf{ngrams}} \\
      & \multicolumn{1}{c}{$0.76 \cdot 10^9$ keys} & \multicolumn{1}{c}{$4.8 \cdot 10^9$ keys} %
      & \multicolumn{1}{c}{$0.76 \cdot 10^9$ keys} & \multicolumn{1}{c}{$7.6 \cdot 10^9$ keys} \\
      \cmidrule(lr){2-3} \cmidrule(lr){4-5}
\textsf{MWHC} & \textbf{2.61} b/key
& \textbf{2.61} b/key
& \textbf{2.61} b/key
& \textbf{2.61} b/key
\\
\textsf{HEM} & 3.16 b/key
& 3.31 b/key
& 3.16 b/key
& 3.05 b/key
\\
\textsf{cmph} & 2.77 b/key
 & \multicolumn{1}{c}{-} & 2.77 b/key
 & \multicolumn{1}{c}{-} \\
      \bottomrule
    \end{tabular}
    \caption{Space comparison of \textsf{MWCH}, \textsf{HEM}, and \textsf{cmph}.}
    \label{tab:space}
  }
\end{table*}

The evaluation of lookup efficiency is much subtler, as it depends on
a number of factors, some of which are subject to hardware
architecture. For this reason, we decided to perform the experiments
on three different machines: an Intel Intel i7-4770 (Haswell) at
3.40GHz, the same Intel i7 (Nehalem) machine used for the construction
experiments (see above), and an AMD Opteron 6276 at 2.3GHz.

For both machines and both datasets we performed lookups of $10$\,M
distinct keys, repeated $10$ times. Since lookup times are in the
order of less than a microsecond, it is impossible to measure
individual lookups accurately; for this reason, we divided the lookups
into $1,\!525$ batches of $2^{16}$ keys each, and measured the average
lookup time for each batch. Out of these average times, we computed
the global average and the standard deviation. The results in
Table~\ref{tab:lookup} show that \textsf{HEM} is slower than
\textsf{MWHC} in all cases. On AMD Opteron the slowdown is the
smallest, ranging from $17\%$ to $20\%$; on the Intel i7 (Nehalem) the
range goes up to $19\%$--$26\%$; on the Intel i7 (Haswell), the most
recent and fastest CPU, the slowdown goes up to $30\%$--$35\%$,
suggesting that as the speed of the CPU increases, the cost of the
causal cache miss caused by the double indirection of \textsf{HEM}
becomes more substantial. In all cases, the standard deviation is
negligibly small, making the comparison statistically significant.

We also remark that our implementation of the \textsf{MWHC} lookup
(which is used also in \textsf{HEM}) is roughly twice as fast than
\textsf{cmph} despite using a sparser ranking table; this is because
to perform the ranking we adopt a \emph{broadword}~\cite{KnuACPBTT}
algorithm that counts the number of non-zero pairs in a $64$-bit words
in just a few non-branching instructions, rather than a linear bit
scan with a loop; the smaller ranking table also imposes a lower cache
pressure. Finally, we use a $64$-bit implementation of the Jenkins
hash function, which is faster on long strings than the $32$-bit one
used in \textsf{cmph}.

\begin{table*}[htbp]
  { \small
    \centering
    \begin{tabular}{@{}lrrrr@{}}\toprule
      & \multicolumn{2}{c}{\textsf{URLs}} %
      & \multicolumn{2}{c}{\textsf{ngrams}} \\
      & \multicolumn{1}{c}{$0.76 \cdot 10^9$ keys} & \multicolumn{1}{c}{$4.8 \cdot 10^9$ keys} %
      & \multicolumn{1}{c}{$0.76 \cdot 10^9$ keys} & \multicolumn{1}{c}{$7.6 \cdot 10^9$ keys} \\
      \cmidrule(lr){2-3} \cmidrule(lr){4-5}
      & \multicolumn{4}{c}{\scriptsize Intel i7 (Haswell)} \\[-2pt] \cmidrule{2-5}
\textsf{MWHC}  & \textbf{219} ns $\pm$ 0.3\%  & \textbf{253} ns $\pm$ 1.3\%  & \textbf{199} ns $\pm$ 0.2\%  & \textbf{251} ns $\pm$ 1.8\% \\
\textsf{HEM}  & 284 ns $\pm$ 0.3\%  & 335 ns $\pm$ 1.1\%  & 262 ns $\pm$ 0.3\%  & 338 ns $\pm$ 0.9\% \\
\textsf{cmph}  & 466 ns $\pm$ 0.3\%  & \multicolumn{1}{c}{-}  & 303 ns $\pm$ 0.3\%  & \multicolumn{1}{c}{-} \\
      & \multicolumn{4}{c}{\scriptsize Intel i7 (Nehalem)} \\[-2pt] \cmidrule{2-5}
\textsf{MWHC}  & \textbf{365} ns $\pm$ 0.1\%  & \textbf{433} ns $\pm$ 0.1\%  & \textbf{334} ns $\pm$ 0.1\%  & \textbf{422} ns $\pm$ 0.2\% \\
\textsf{HEM}  & 450 ns $\pm$ 0.1\%  & 523 ns $\pm$ 0.1\%  & 420 ns $\pm$ 0.1\%  & 502 ns $\pm$ 0.7\% \\
\textsf{cmph}  & 799 ns $\pm$ 0.1\%  & \multicolumn{1}{c}{-}  & 532 ns $\pm$ 0.1\%  & \multicolumn{1}{c}{-} \\
      & \multicolumn{4}{c}{\scriptsize AMD Opteron} \\[-2pt] \cmidrule{2-5}
\textsf{MWHC}  & \textbf{415} ns $\pm$ 0.1\%  & \textbf{419} ns $\pm$ 0.1\%  & \textbf{373} ns $\pm$ 0.1\%  & \textbf{386} ns $\pm$ 0.1\% \\
\textsf{HEM}  & 484 ns $\pm$ 0.1\%  & 493 ns $\pm$ 0.1\%  & 442 ns $\pm$ 0.2\%  & 463 ns $\pm$ 0.1\% \\
\textsf{cmph}  & 908 ns $\pm$ 0.2\%  & \multicolumn{1}{c}{-}  & 578 ns $\pm$ 0.3\%  & \multicolumn{1}{c}{-} \\
      \bottomrule
    \end{tabular}
    \caption{Lookup-time comparison (with relative standard deviation)
      of \textsf{MWCH}, \textsf{HEM}, and \textsf{cmph}.}
    \label{tab:lookup}
  }
\end{table*}

\bibliographystyle{plain}
\bibliography{cofun_arxiv}

\end{document}